# An Implementation Approach and Performance Analysis of Image Sensor Based Multilateral Indoor Localization and Navigation System

Md. Shahjalal, Md. Tanvir Hossan, Moh. Khalid Hasan, Mostafa Zaman Chowdhury,
Nam Tuan Le, and Yeong Min Jang
Department of Electronics Engineering, Kookmin University, Seoul, Korea
Email: mdshahjalal26@ieee.org, yjang@kookmin.ac.kr

**Abstract-** Optical camera communication (OCC) exhibits considerable importance nowadays in various indoor camera based services such as smart home and robot-based automation. An android smart phone camera that is mounted on a mobile robot (MR) offers a uniform communication distance when the camera remains at the same level that can reduce the communication error rate. Indoor mobile robot navigation (MRN) is considered to be a promising OCC application in which the white light emitting diodes (LEDs) and an MR camera are used as transmitters and receiver respectively. Positioning is a key issue in MRN systems in terms of accuracy, data rate, and distance. We propose an indoor navigation and positioning combined algorithm and further evaluate its performance. An android application is developed to support data acquisition from multiple simultaneous transmitter links. Experimentally, we received data from four links which are required to ensure a higher positioning accuracy.

**Keywords**- Image sensor, optical camera communication, distance measurement, indoor positioning.

## 1. Introduction

Currently, optical wireless communication (OWC) is extensively used to mitigate the data traffic from mobile communications among which the ultraviolet (UV), infrared (IR), or visible light (VL) spectrums are used as the propagation mediums. The VL spectrum is extensively used by OCC, light fidelity (Li-Fi), and visible light communication (VLC) [1], [2]. OCC is becoming a promising technology in terms of localization and navigation because global positioning system (GPS) using radio frequency (RF) based wireless networks such as wireless fidelity (Wi-Fi), ultra wideband (UWB), Bluetooth, wireless local area network (WLAN) which are not much accurate location based communication services. Also GPS works well only in the case of outdoor applications, such as tracking and robot navigation, and it is considerably difficult to estimate the accurate locations using RF technology in an indoor environment. WLAN based localization schemes such as AOA, TDOA, WLAN fingerprint, have several challenges as it was developed initially only for wireless networking. In the AOA and TDOA approaches the user have to know the locations of access points (APs) and are not able to provide maximum network coverage. Also the localization errors are high for the NLOS cases [3]. Although fingerprint method which uses power of the received signal have recently captures lots of attention, it faces several challenges. Specific RSS measurement is not possible for individual APs due to dense deployment. Moreover, surveying stage is more time consuming and records the RSS for a time period which increases the complexity of the system. Also RSS weakens in case for NLOS propagation due to the presence of obstacles like walls, furniture's and doors [4]. In addition, if there are many WLANs in a typical region would cause extra interference between each other. Whereas, OCC is a new technology offers low cost, simple, low latency communication and higher positioning accuracy. We used smartphone based localization technique using OCC where image is processed by controlling the camera properties so that other lights cannot make any interference among each other. As each room contains LED light so there is no possibility that objects will make any interruption during the LOS communication.

OCC uses an unlicensed spectrum in which the LED infrastructures are used as the transmitters. LED is an electronic light source in which illumination intensity can be controlled at a high frequency using an externally driven circuit. LED is a useful transmitter because it is energy efficient, common indoor lighting infrastructures, cheap; additionally, a highly accurate data transmission is possible due to the variable luminary properties of LED. Digital data is transmitted through the LED light by varying the properties of LED according to the different modulation schemes. Typically, for low frame rate (25~30 fps) commercial cameras under-sampling modulation techniques are used. This technique includes phase-shift on-off keying (PSOOK) [5], frequency-shift on-off keying (FSOOK) [6], m-array pulse amplitude modulation (PAM) [7] and so on.

A color camera, which is used as the receiver for OCC applications, typically comprises an image sensor (IS), lens, and Bayer filter. A camera exhibits several benefits over a photo diode (PD) in terms of the extended field of view (FOV) and because of the fact that the pixels of the IS can receive light from various directions, thereby providing a high signal to noise ratio (SNR). Localization is possible at the cm-level of accuracy in an indoor environment under stable illumination from an external light source. Several reports on positioning using a VLC system have already been proposed. In one report, a VLC-based positioning system in which the LEDs wirelessly transmit the location information has been proposed [8]. The location information signals were received by a PD, and the spatial distribution was measured at each reference point. Further, the information from people or objects can be obtained using map positioning. There are some localization or navigation services using OCC exhibit an external cloud server to store the mapping data and to compute the position at which the camera is connected to server via Wi-Fi or Bluetooth during location estimation [9]. This approach exhibits a few drawbacks because it requires an external server and an additional RF network, which is expensive and time consuming.

For localizing a rolling shutter (RS) camera that was mounted on an MR, the location information should be received using the camera from multiple LEDs. We propose an android phone mounted mobile robot positioning (MRP) approach using multiple frequency-shift keying (MFSK) in which we are able to obtain four locations ID from four different LEDs simultaneously. We enhance distance measurement technique that allows maximum 2 cm error at 100 cm horizontal distance in case of single LED inside the FOV when camera and LED are not even in vertically line-of-sight (LOS). An MRN algorithm is also proposed to navigate MR to different location. Relevant works on indoor localization is described in Section 2. Section 3 describes the overall scenario of the indoor MRP and MRN systems as well as the transmitter characteristics. A brief description of the android application and the openCV libraries, rolling shutter effects, and MR specifications are provided in Section 4. The proposed algorithm and the remaining operational characteristics, such as the camera FOV and the exploration of ID distribution systems, are described in Section 5. In Section 6, the details of the accuracy enhancement technique are provided. A demonstration scenario and results are presented in Section 7. Finally, this research work concludes in Section 8.

## 2. Related Works

Currently, camera-bearing mobile phones are commonly used and are important commodities in e-commerce. Therefore, location based services (LBS) have become an important issue. To ensure better performance in LBS, it is mandatory to measure the location of these mobile devices accurately especially indoors.

GPS is a pseudo-lite system, which faces challenges at the point of indoor localization. This system is a LOS based localization solution which accumulates the sensor information from satellites that are located at a considerable distance (20,000 km) from the ground. The signals of the satellites are interrupted by the obstacles on the ground, such as trees and buildings, due to the LOS channel characteristic and the considerable distance between satellites and sensors. Therefore, a considerably large modification is required to make the GPS system to be suitable for indoor localization. For example, the integration of GPS signals with the indoor transmitting antennas has been reported to localize sensor nodes [10]; however, it is not cost effective, and the localization error is observed to be 1 m in delay of 10 sec.

The distinctness and novel characteristics of OWC-based techniques are considered important supporting candidate over existing solutions for indoor localization and navigation scenarios. Several approaches of VL-based localization and navigation schemes for mobiles, PDs, or wearable computers have been reported [11]-[17]. There is still considerable debate about the value of localization resolution. A lower value of localization resolution was observed from the simulation results of various conditional approaches. An important subpart of OWC is OCC in which the camera receives a signal from the modulated light beam that was observed from an LED. An IS detects the intensity of the metameric LEDs within a fixed indoor environment [18]. The localization performance exhibits a 1-degree orientation measurement, and the calculated resolution was observed to be 1.5 and 3.58 cm for 2D and 3D space, respectively. Without measuring the angular displacement, 0.001-m localization resolution has been observed [19]. In a similar approach, the localization using an LED transmitter and a camera as the receiver has been discussed by other groups [20], [21]. Additionally, the information from an accelerometer sensor has been included; further, the demodulated data from the IS could be used to improve the overall performance in a 3D indoor environment [22].

Popular methods for localization can be used to measure the signal strength of the receiver, i.e., a photodiode or camera. This signal can be either visible light or RF. An RSS-based localization scheme for an aperture-based receiver, which exhibits a wide FOV and a better angular diversity, has been proposed [23]. They derived the Cramer-Rao lower bound of position estimation for improving the overall performance. Meanwhile, a generic framework of RSS has been introduced to enrich the positioning performance [24]. Alternatively, the combination of spatiotemporal constraints in RSS fingerprint-based localization has achieved the same purpose [25]. Compressive sensing has been applied to improve the localization performance in noisy scenarios [26]. The implication of the RSS approach was also observed in the VL-based positioning schemes [27]-[29]. Here, the localization resolution was observed to be 4 m [30]. Importantly, an artificial intelligence algorithm that was applied to collect the location-labeled fingerprints was reported for optimizing the overall performance [31], [32]. Although the positioning accuracy that was obtained for the RSS fingerprints was observed to be 80% [33], a 3D space fingerprint required considerable overhead for positioning a high speed object as compared with that obtained in a simpler 2D space fingerprint.

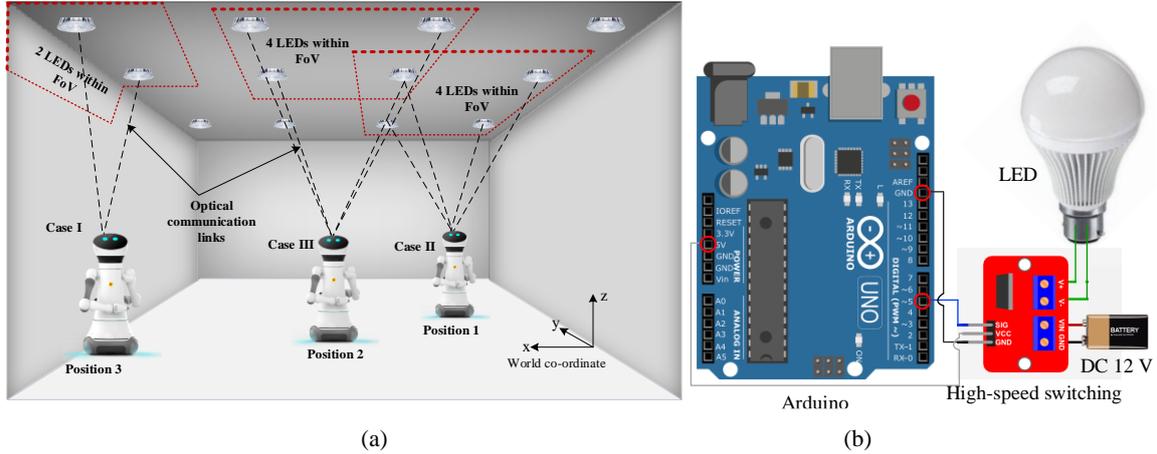

Figure 1. A scenario of the proposed model (a) MRP and MRN system using OCC, (b) Transmitter set.

Time-of-arrival (TOA) and time-difference-of-arrival (TDOA) are two methods that have been used several times to provide a solution for localization in indoor environments. A TDOA-based localization scheme has been proposed in which a light beam with a unique frequency was transmitted in different phases [34]- [36]. A TOA-based localization algorithm has also been developed [37]. The most important issue is that the TOA- and TDOA-based localization schemes are cost-effective and accurate. Furthermore, these schemes depend on the location information from a central node as well as from other reference nodes in the same indoor space. An extended Kalman filter-based TDOA approach that ignores the impact of this dependency on the reference node information has also been proposed [38]. However, deploying such a tracking algorithm is not always advantageous because the extended Kalman filter failed to accurately estimate the position as a first order approximation.

Angle-of-arrival (AOA) is another method that can be applied for indoor localization. In this approach, the receiver estimates the angle of the transmitted signal by calculating the time difference of arrival from the individual transmitter. Transmitting the gain difference from the LED has been considered by AOA for indoor localization [30], [39], [40]. Simulation results depicted that the average value for localization resolution was 3.5 cm. However, in our proposed system, it is not essential to gather the angle information from LEDs to measure the position of the camera. The position is calculated with the help of distance comparison among several LEDs (e.g., more than two LEDs). These distances are calculated using the photogrammetry technique [43], where the occupied image area of the LEDs on the image sensor varies with the distance. Furthermore, AOA exhibits some disadvantages, e.g., the accuracy degrades with the increasing distance between the transmitter and receiver, the reflections of the signal from the multipath adds some error during the location measurement, requires large and complex hardwire devices as well as the shadowing and directivity of the measuring aperture have significant effect on the overall performance of location estimation.

## 3. Indoor Transmitter Scenario

Figure 1(a) depicts a scenario of the proposed indoor transmitter and receiver scenario for MR positioning and navigation system using OCC. A circular shaped white LED having a diameter of 9.5 cm was used as the transmitter. Because the LED exhibits an altering luminary property, it can flicker at ultrahigh speeds that are beyond the perception of the human eye. The indoor location ID is modulated with a high modulation frequency and encoded with additional bits using an LED driver circuit. This encoded signal is further fed to the LED, which continuously transmits this ID using the flickering light. The driver circuit comprises a 12-V dc power supply, an Arduino board, and a high speed switching device including metal-oxide-semiconductor field-effect transistor (MOSFET). The transmitter set with an LED driver circuit is depicted in Figure 1(b). LEDs are installed on the ceiling, and the location number is transmitted to the MR receiver camera. In our system, transmitters are arranged in a square-shaped regular pattern in which each side of LEDs are 50 cm apart from each other. In this case, the positioning system of the camera is mounted on top of the MR, which maintains an equal height along the z-axis. Therefore, the LEDs transmit the modulated data bits that contain only the (x, y) location information.

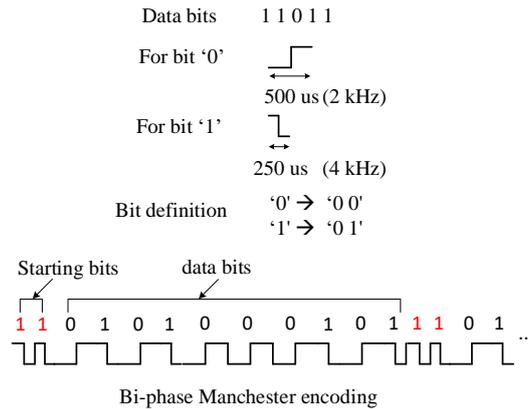

Figure 2. Transmitted data bit mechanisms.

The visual flickering in an optical wireless communication is a major problem that can be defined as the change in the illumination of the LED when it transmits a binary '1,' '0' code-word in terms of the light that is perceived by the human eyes. Generally, human eyes can avoid flickering if the data is modulated with a frequency that is greater than 200 Hz [41]. The LED transmitter should be driven using the accurate LED driver circuit to avoid any noticeable flicker. Multiple frequency-shift keying (MFSK) is used as the modulation technique to generate the modulated binary code-words using the LED driver circuit. Figure 2 depicts the characteristics of the transmitted bit pattern. Our proposed study uses the modulation frequencies 2 and 4 kHz, which are safe enough to avoid flickering. Further, the modulated signal can be encoded using bi-phase Manchester line coding. Each of the binary bits, 1 and 0, is defined using a different code. Bit '1' is defined as '01,' whereas bit '0' is defined as '00.' We transmitted five bits of data from each LED. Therefore, in a single indoor system, 32 location IDs can be provided to 32 LEDs. Furthermore, if the actual data is '11011,' the encoded signal will be '0101000101' after bi-phase Manchester line coding as shown in Figure 2. Although there are 3 consecutive '0's in this example, the bi-phase Manchester encoding eliminates the possibility of visual flickering because it confirms one transition in the middle of each bit. As this code stream is continuously transmitted, a starting bit symbol '11' is added at the beginning of the code-word to differentiate between every simultaneous code-word.

## 4. Receiver Characteristics

*4.1 Android application background.* An application is designed for the android studio platform using the camera 2 hardware package which is added in API level 21. Some android features are described in this study that are used to create the camera device and process the requests. One request acts to capture a single frame and output a set of image buffers for the request. A *cameraManager* is required when multiple requests are in queue to maintain the full frame rate. To accumulate the captured frame data a *getCameraCharacteristics* class is used. The image data is encapsulated in *Image* objects and can be directly accessed through a class named *ImageReader* using a YUV_420_888 format. The image data is rendered onto a surface with defined size. For processing the captured image frames openCV320 libraries were imported into the application. Initially, the camera shutter speed is controlled to focus only to the white LEDs and to detect only the region of interest (RoI).

*4.2 Rolling shutter effect.* Currently, most of the consumer cameras contain complementary metal–oxide–semiconductor (CMOS) sensors, which include rolling shutter mechanisms. The sequential read-out technique is a key feature of the rolling shutter camera in which each frame is not captured at exactly the same instant [42]. The scene is captured by scanning rapidly, either horizontally or vertically, unlike a global shutter camera in which the complete scene is captured at the same time. In a rolling shutter camera, each row of the pixels is exposed at once at the exposure time. The read-out time protects the rows of pixels from overlapping. In a single captured image, the rolling shutter allows multiple exposures. Therefore, for an LED, which flickers on-off according to the modulated binary bit stream, the captured image contains a bunch of dark and white strips. The width of the strips depends on the modulation frequencies, and the number of strips depends on the distance.

Although a different camera receives signals of similar frequencies from the same transmitter, the width of the strips is different as the specifications of the camera sensors may differ from device to device. In our system, a multiple frequency-modulated signal is transmitted, and the dark and white strips are received at different distances, as depicted in Figure 3. The camera captured the pictures with a fixed clockwise rotation of *ImageView* of $270°$. As the distance increases, the size of the LED in the IS decreases; further, the number of strips also decreases because the width remains identical for fixed frequencies at any distance for a particular camera.

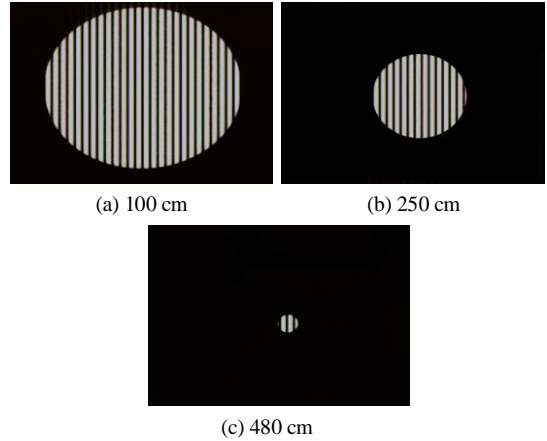

(a) 100 cm  (b) 250 cm

(c) 480 cm

Figure 3. Experimentally, the number of dark and white strips decreases with the increase in distance. (a) 52 strips at a distance of 100 cm (b) 28 strips at a distance of 250 cm. (c) 5 strips at a distance of 480 cm.

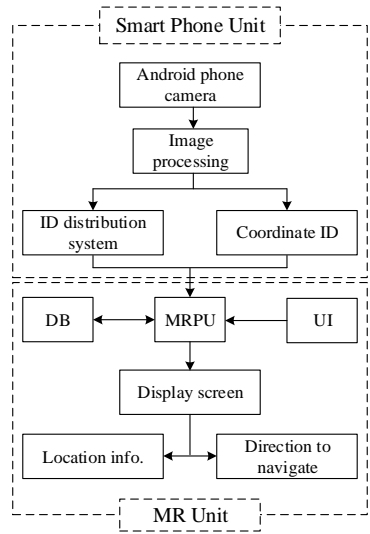

Figure 4. Block diagram of the mobile robot functionalities.

*4.3 Introduction to MR functionalities.* Localization and navigation are challenging issues in optical camera communication for the existing MRs. The initial requirement is to connect to the input interface of the MR to provide necessary data from the android phone such as the location ID and its distribution system. When a camera captures the rolling images from multiple LEDs, it processes the images and obtains the locations, thereby making a decision about the location distribution system and then feeds the data to the MR. The MR must exhibit a user input interface (UI) to set a target location ID, a data base (DB) to store the ID distribution information, and a display screen to exhibit the location information and the subsequent direction to move after making a decision. Figure 4 depicts the specifications of the MR and its required functionalities. It also exhibits a data and image-processing unit to compare the IDs and a database to store the previous tracking information. The MR receives the IDs and can therefore localize itself. It also has a capability to compare the stored IDs and navigate by itself if the user sets a destination.

## 5. Operational Principle

*5.1 Navigation algorithm.* The objective of the MRN system is to navigate an MR by receiving the locations or location ID using multiple LED lights. Figure 5 describes a flowchart of the applied navigation algorithm in our proposed system. A user inputs a target location into the smart phone that is mounted on the top of an indoor MR. Because the complete environment in the room is unknown to an MR, it should fix its position first, which indicates that a location should be observed inside the room. Therefore, it estimates the number of LEDs that are within the camera's FOV. To perform navigation, the MR should perceive the direction in which the location values are changing. Thus, a minimum of three LEDs should be within the camera's FOV, as depicted in Figure 1(a) in previous section. Here, in case I, the IS can only detect two LEDs among which each LED exhibits the same value for the x location; further, it is difficult to observe a variation in the value of the x location. In case II and III, there are four LEDs within the camera's FOV as MR is situated deep inside the room. If three or more LEDs are detected using the IS, the smart phone receives the LED locations and compares them with the target location. If the target location matches with any of the received locations, the MR moves in a horizontal distance toward the LED that is located directly below the target location.

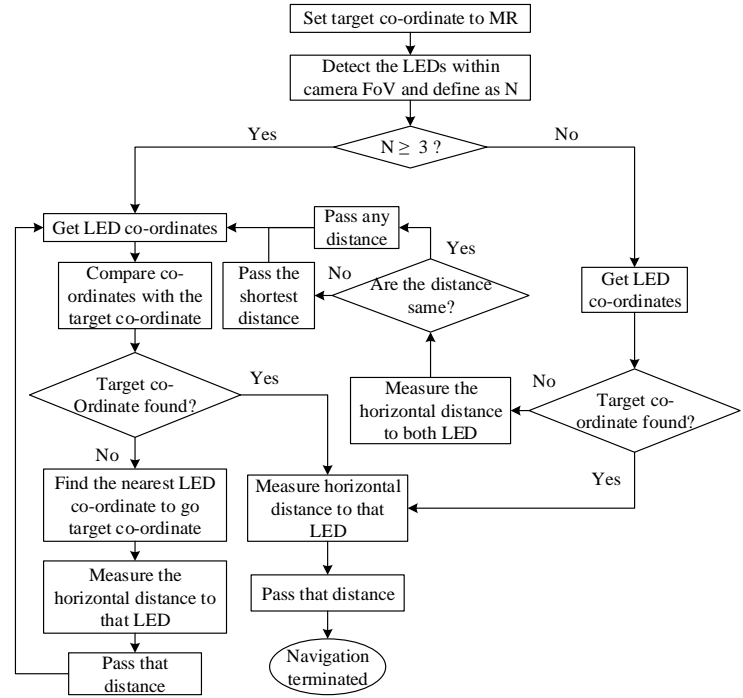

Figure 5. The mobile robot navigation flowchart.

As mentioned in the previous paragraph, to localize the MR to a floor position of a target LED, we need to measure the horizontal distance from the MR to that floor position. Here, floor position of an LED means the position on the floor which is just below the LED. Suppose, the target LED is transmitting (x, y) location ID. When the MR is located in its initial position as depicts in Figure 6, the implemented android smartphone app calculates the direct distance, D from the ceiling LED. The direction towards the LED is predicted from the image formation on the IS (image formation technique on IS is well described in [41]). After obtaining the direct distance, we can easily find the horizontal distance, s using Pythagoras distance formula for right angle triangle as the vertical distance is fixed inside a room. After calculating the value of s and understanding the direction to move, MR navigates this distance and fixes its location to (x, y) location. If the target location is not matched with any of the received location ID, the MR estimates the nearest LED that is required to reach the destination. Subsequently, the MR traverses a certain amount of distance to go a point that is exactly below that particular LED. Further, the MR again receives the locations of the LEDs that are within camera's FOV. Subsequently, the procedure continues until the destination is reached. If the camera detects two LEDs within its FOV, the MR cannot detect the location distribution system. Therefore, it initially chooses one ID and goes to a point that is located exactly below that LED, and then MR continues with the aforementioned procedure.

*5.2 Camera field of view.* We considered the indoor environment to possess a uniform square distribution of LEDs. Figure 7 depicts the manner in which the number of LEDs within camera's FOV depends on different factors such as the height of the room, the distance between two adjacent LEDs, and the angle of view (AOV) i.e., the FOV angle. For the purpose of navigation, the MR should understand the location distribution system among the LEDs. It is obvious that at least two diagonally located LED locations should be considered to make a decision about the direction in which the X and Y locations are increasing or decreasing. A relation has been derived to estimate the number of LEDs that were captured within the FOV at any time considering the camera view angle, the height of the roof, the surface area taken by a single LED.

The horizontal and vertical FOV can be estimated using the following two equations:

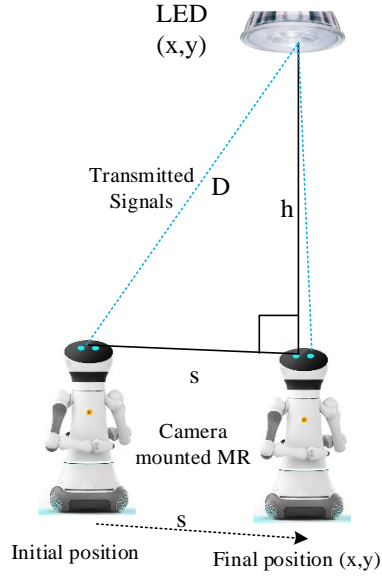

Figure 6. Calculation procedure of the distance to move horizontally.

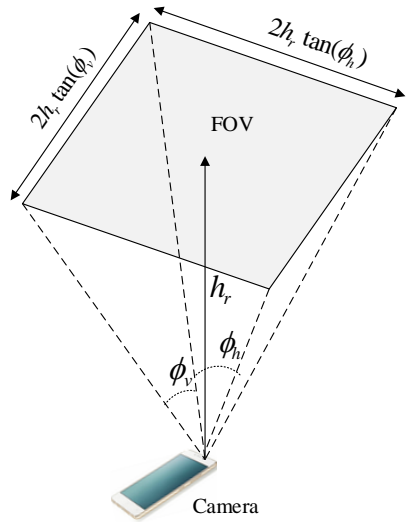

Figure 7. Analysis of the camera's FOV.

$$\phi_h = 2\tan^{-1}\frac{d_h}{2f} \quad (1)$$

$$\phi_v = 2\tan^{-1}\frac{d_v}{2f} \quad (2)$$

where, $\phi_v$ and $\phi_h$ is the FOV in vertical and horizontal directions and $d_h$, $d_v$ are the sensor dimensions respectively. $f$ represents the focal length of the camera lens. Figure 7 shows the camera FOV and its area is projected vertically on the roof of height $h_r$. We can represent the area of the FOV using the following equation.

$$A_{FOV} = 4h_r^2 \tan(\phi_v)\tan(\phi_h) \quad (3)$$

Because the LEDs are distributed in a square pattern and each LED separated by a distance $a$ along each side (where, $a >$ LED diameter) then each LED will occupy an area of $a^2$.

Now we can find the number of LED will be in the camera FOV using the above (3)

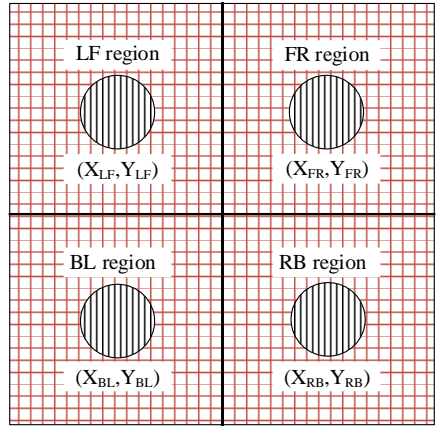

Figure 8. Region basis locations selection and exploring their distribution system.

$$N_{LED} = \frac{4h_r^2 \tan(\phi_v)\tan(\phi_h)}{a^2} \quad (4)$$

*5.3 Exploring the Id distribution system.* The smart phone that was used to perform this study was a Samsung S7 edge, which contained a 26-mm camera and a focal length of 4.2 mm. The height of the room that was observed using the camera was 2.56 m. We set up the android camera with a height and width of 600 and 800 px, respectively. In a general scenario, there were always at least four LEDs within the camera's FOV under this type of consideration. To define the coordination distribution system, we divided the IS into four regions, i.e., left-front region (LFR), front-right region (FRR), right-back region (RBR), and back-left region (BLR), as depicted in Figure 8, where each region contains an equal number of pixels. The camera will detect a single LED per region and receive the location ID as well. We defined the locations on a regional basis, which are $(X_{LF}, Y_{LF})$, $(X_{FR}, Y_{FR})$, $(X_{RB}, Y_{RB})$, and $(X_{BL}, Y_{BL})$, respectively, within the four regions. It is obvious that the LEDs will be present either in both the LF or RB regions or in both the FR and BL regions. If $X_{LF} < X_{RB}$ or $X_{FR} > X_{BL}$, the x-location increases in the right direction; otherwise, it increases in the left direction. Further, if $Y_{LF} > Y_{RB}$ or $Y_{FR} > Y_{BL}$, the y-location increases in the front direction; otherwise, it increases in the back direction.

## 6. Performance analysis

*6.1 Measurement of distance accuracy.* The direct distance from the camera to an LED can be measured by considering a few distance parameters. If $D$ is the direct distance of the LED $(x_i, y_j)$ from the camera lens and $d$ is the distance of the formed image on the sensor from the camera lens, then we can write from the general lens equation:

$$\frac{1}{D} + \frac{1}{d} = \frac{1}{F} \quad (5)$$

$$\frac{d}{D} = \frac{F}{D-F} \quad (6)$$

where, F is the focal length of the lens. Here, magnification factor is required to introduce and compare the relative size of the actual size and the detected size on the IS.

The magnification factor, m can be written as follows:

$$m = \sqrt{\frac{a_{ij}}{A_{ij}}} = \frac{d}{D} \quad (7)$$

where, $A_{ij}$ is the actual surface are of the nearest LED $(x_i, y_j)$ and the area covered by the detected image on the IS is $a_{ij}$.

Normally, it is observed that F<<D, therefore, we can combine the above equation as follows:

$$a_{ij} = m^2 A_{ij} \quad (8)$$

Table 1. Distance measurement data from an LED at different floor positions.

| Horizontal distance, $d_h$ (cm) | Measured direct distance, (cm) | Actual direct distance, (cm) | Error in distance measurement (cm) | Detected area (px) | 1st Radius, $r$ (px) | 2nd Radius, $r'$ (px) |
|---|---|---|---|---|---|---|
| 00 | 255.5 | 255.5 | 0 | 348 | | 11 |
| 10 | 255.88 | 255.9 | 0.02 | 347 | | 10 |
| 20 | 256.65 | 256.7 | 0.05 | 341 | | 10 |
| 30 | 257.6 | 257.7 | 0.1 | 337 | | 10 |
| 40 | 258.85 | 259 | 0.15 | 336 | | 10 |
| 50 | 260 | 260.2 | 0.2 | 330 | 11 | 9 |
| 60 | 262.5 | 262.8 | 0.3 | 324 | | 9 |
| 70 | 264.3 | 264.7 | 0.4 | 314 | | 9 |
| 80 | 267.1 | 267.7 | 0.6 | 306 | | 9 |
| 90 | 269.5 | 270.61 | 1.11 | 298 | | 8 |
| 100 | 272.5 | 274.5 | 2 | 288 | | 8 |

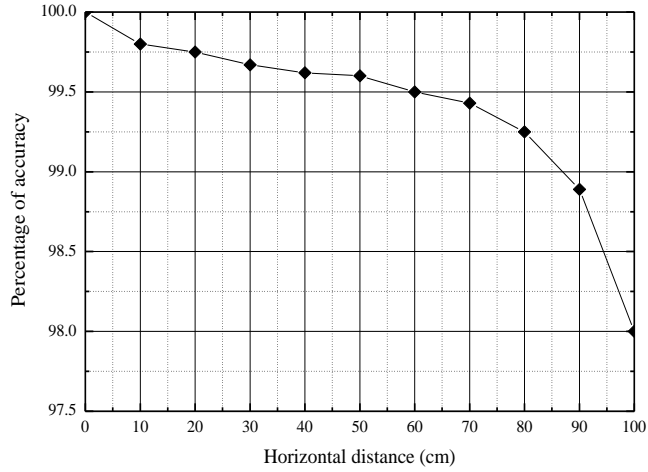

Figure 9. Percentage of accuracy of the measured direct distances from the camera to the LED at different horizontal points.

$$a_{ij} = \frac{F^2}{D^2} A_{ij} \qquad (9)$$

$$D = F\sqrt{\frac{A_{ij}}{a_{ij}}} \qquad (10)$$

where, the focal length F and the actual LED size $A_{ij}$ are known. The only requirement is to obtain the detected LED area, $a_{ij}$ from the IS.

Using (10), the distance can be measured accurately at a position at which the camera is vertically in the LOS of the LED. But when it replaces to some other places which are not just below the LED experiences some error due to virtual change of the actual size of the circular LED to eclipse. This happens due to the radial distortion of the camera [44]. The wide-angle camera, short focal-lengths camera or the fisheye camera are the main reason behind the radial distortion. Determining intrinsic, extrinsic and distortion coefficient of IS and lens of the camera is required to omit the effect of radial distortion. This radial distortion makes a significant transformation of the projected object from the actual object with a normal focal length. To verify the localization scheme, a back-projection method deploys to measure area of LED from distorted image from read-world space [45]. In the following section we will exhibit a figure to illustrate the accuracy of the distance measurement. Although at the boundary position the accuracy remains greater than 98%, we have undertaken one further step to mitigate this error. Let's take $r$ as the radius of circular LED and the area will be $\pi r^2$. Now when circular shape changes to eclipse, one radius $r$ reduces to $r'$ and the area becomes $\pi r r'$. Table 1, presents the data measured from a LED with a size of 71 cm$^2$ located at a ceiling of height of 2.56 m. The pixels allocated for the detected LED at different positions are recorded to show at far distance detected contour becomes smaller than nearer. We tabulated the radius of eclipse which is required to calculate the actual size of eclipse and correct the positioning error.

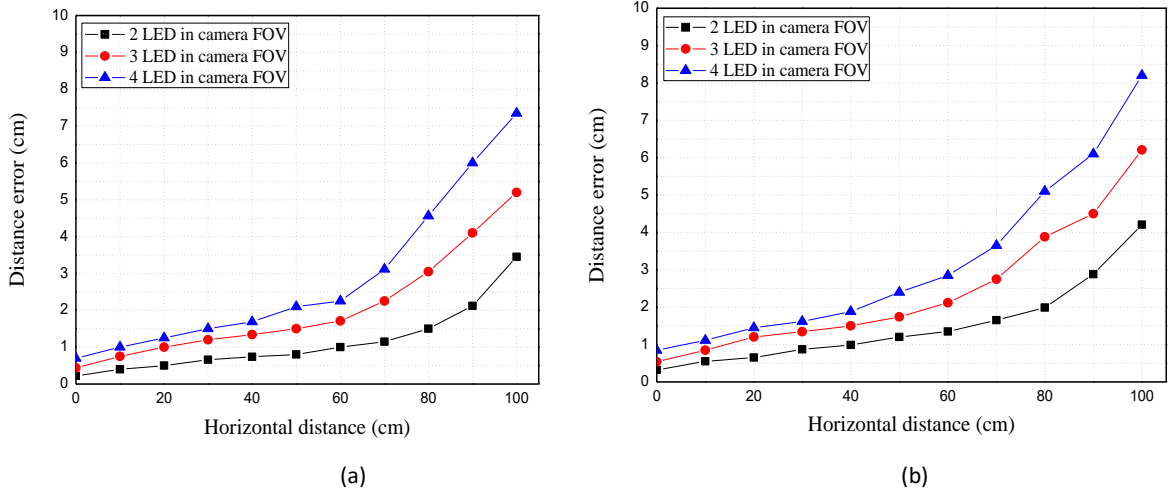

Figure 10. Distance error comparison in terms of variation of no. of LEDs inside the camera FOV at different horizontal points. (a) inter LED distance = 50 cm, (b) inter LED distance = 100 cm.

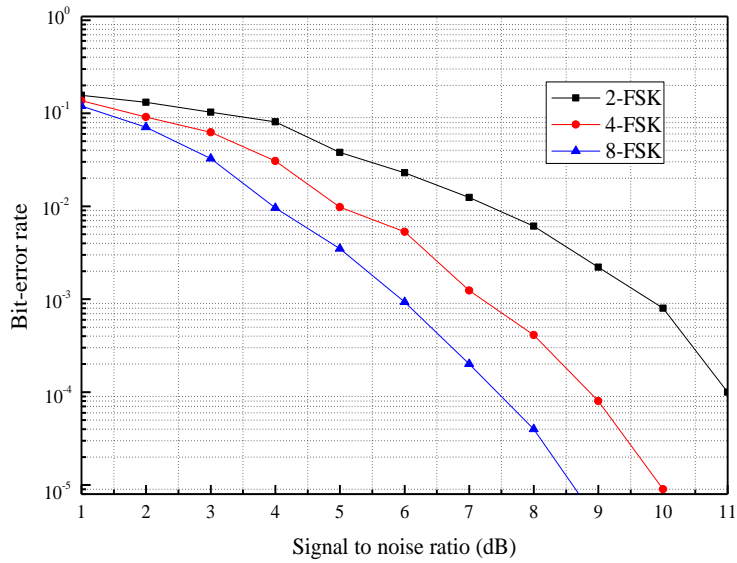

Figure 11. BER comparison for M-FSK (M=2,4,8) in indoor wireless Rician fading channel.

This radius data is integrated within the application and will be called when this radius will be measured by smartphone through image processing. Comparing the existing data with the actual area of the LED can be predict to measure the direct distance.

The distance between the LED and the camera changes when the MR moves horizontally. The direct distance from a particular LED is required to calculate the horizontal distance that is required to move and to get the floor location of that particular LED. We used (10) to measure the direct distance between the camera and the LED. We graphically represented the percentage of accuracy that is observed while measuring the direct distance at a particular point on the floor in Figure 9. The distance is measured with 100% accuracy when the camera is situated directly below the LED. When the camera is shifted horizontally far away from the floor location of the LED, the accuracy is reduced.

To evaluate the performance of the proposed localization algorithm we have tested the accuracy of measuring the distance for different type of scenarios. We have got the distance measurement error of 2 cm maximum in case of single LED inside the camera FOV. However, when there multiple LEDs inside the camera FOV, the distance error increases as the probability of LED being at the middle of the IS reduces. Figure 10 depicts the comparison of error in measuring the distance between LEDs and camera for different cases varying the no. of LEDs into the camera FOV. We have taken two scenarios first is for the case where the inter LED distance is 50 cm and second is for the case where that is 100 cm. In each scenario we have considered 2 to 4 LEDs for the comparison. We started measuring from the floor position of an LED and took 10 more reading going away about 10 cm in each step. We can see from the figure that if there are more LEDs into the IS at the time the error for measuring the distance increases because the chance of an LED to be at the center of IS reduces. As our system comprises of LEDs those are distributed in a rectangular pattern there are more possibility of being 4 LEDs at a time in the IS. And for this case we have to consider up to 7.5 cm distance error for positioning and navigation purpose when the inter LED distance is 50 cm. But for the second case described in Figure 10 (b) the value of the distance error increases up to 8.5 cm for 4 LEDs due to increase of the inter LED spacing to 100 cm.

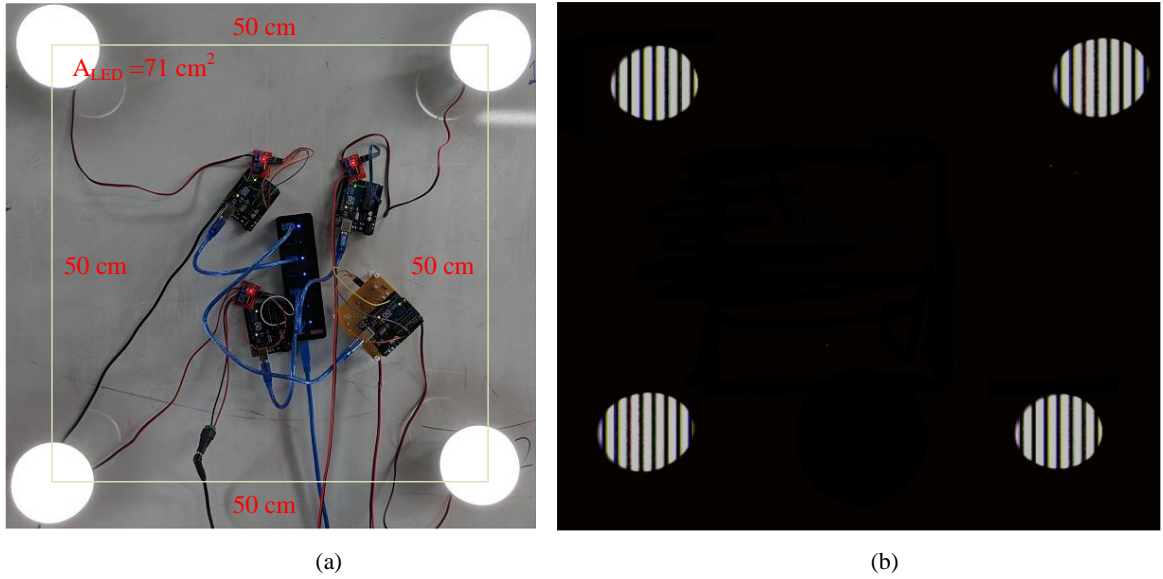

(a) (b)

Figure 12. Experimental setup. (a) four LEDs that are 50 cm apart from each other transmit the location IDs. (b) Captured image after setting the exposure time to 1/8000 sec.

Though the distance error reducing with the reduced no. of LEDs being allowed into the camera FOV, it is required to have more than two LEDs into the camera FOV to get the navigation information and we should accept this error anyway. Also, we have described about a way in this section that can be applied to reduce the distance error to a certain level for moving to a horizontal distance to reach the target LED location increasing the positioning performance.

*6.2 Bit error rate for MFSK.* To evaluate bit error rate (BER) performance we considered an indoor wireless Rician fading channel. MFSK is simple and flexible for multi-link communication over this channel. It is known [46] that the bit error probability for MFSK is:

$$\rho_s = \frac{1}{M}\sum_{i=2}^{M}(-1)^i \binom{M}{i}\exp[-(1-\frac{1}{i})\rho] \tag{11}$$

where, M is the number of order and $\rho$ is the instantaneous signal-to-noise ratio (SNR) value. In Figure 11, we depicted a BER comparison graph between different modulation orders (M=2,4,8). We have seen for increasing M, bit error probability decreases. When the transmission is done keeping same bit energy which means that the symbol duration becomes double for the case of 4-FSK than 2-FSK, maintaining the same bit rate meanwhile. In such case the BER becomes lower as symbol duration increases with higher modulation index.

## 7. Demonstration results

We used a platform to demonstrate our OCC system to evaluate its performance. To understand the ID distribution system, we should receive the location information using a minimum of two diagonally located LEDs that are described in section V. The platform comprises four LED transmitter links that are placed in a square shape, as depicted in Figure 12(a). The LEDs are arranged by maintaining a distance of 50 cm between them. For better explanation, we considered these transmitters to be a single unit. These four transmitters are linked continuously and transmit four different locations, which are encoded with the help of a PC and an LED driver circuit.

To focus only on the target-transmitting LED, we controlled the camera exposure time. The exposure time is the duration at which each pixel is exposed to light. Using this process, we can obtain the dark and white strips at a region at which the LEDs are captured by darkening all the parts of the image except the parts that depict the target LEDs. Figure 12(b) depicts the captured image using the Samsung S7 (edge) camera of the transmitting LED unit after setting the camera's *CaptureRequest.SENSORE_EXPOSURE_TIME* to 1/8000 sec. We can observe that all the four LEDs represent same dark and bright strip width. This is because same modulation frequencies were used for all the four transmitters, and the frequencies ranged from 2 to 4 kHz. Generally, the characteristics of the rolling image are such that the strips can be formed horizontally. However, in our case, we rotated the image using *imageView* at 270°, and this is the reason for the observation of vertically formed strips.

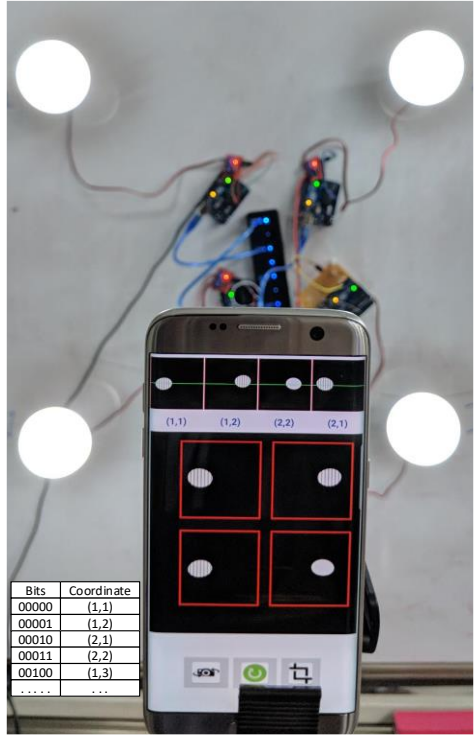

Fig 13. Implemented android application receiving multiple LEDs ID simultaneously.

To demodulate the data from the detected strip pattern we should initially measure the width of each dark and white strip. We know different camera has different frame rate and also for a fixed camera, frame rate fluctuates within a short range. For this reason, the width also varies and as the width is measured in a scale of number of pixels in a row, it should be an integer value. In our system, we observed brighter strips at which the width varied between 5 and 7 pixels, whereas it varied between 1 and 3 pixels for darker strips. Therefore, the mid-value was considered to be four to distinguish between dark and white patterns. When the measured width was greater than four, smart phone will consider the LED to be in the ON state and store a binary '1' value. Otherwise, the LED is assumed to be in the OFF state and stores a '0' value. Figure 13 illustrates the android implementation software that can simultaneously receive data from four LEDs. We received five bits of ID per LED and defined each combination of the received binary ID to a fixed location in the application system. The table in Figure 13 depicts the manner in which the location is assigned to different IDs. To receive multiple LEDs, the overall time for processing one frame should be within the critical frame time. Otherwise the application may unfortunately be stopped. The critical frame time is defined as the maximum processing time for one frame. In our system, we used 20 fps to limit the critical frame time to 50 msec. We add a starting symbol to recognize when the data is starting. So, the time to get the data obviously depends on the timing of getting the frame containing that symbol. We test in real-time debugging the software in android studio. We observe the time which is required for initialization, frame processing, demodulating, and device response. We figure out that almost 1000 ms is required to debug and getting device response when communicating with 4 LEDs simultaneously. As each LED is transmitting 10 bits, we got data at 40 bps rate for 4 LEDs with this device.

## 8. Conclusion

This work focuses on the reception of data from multiple indoor LED transmitters as well as on the localization and navigation of the MR, processing the received data. The indoor LED locations are received as binary data of five bits per link at a data rate of 40 bps. The proposed algorithm was implemented in an indoor platform where four LEDs having sizes of 71 cm$^2$ were kept at a distance of 50 cm from each other. The four LEDs transmit four different IDs, and an android application that was developed to receive these IDs performed well simultaneously. We were able to communicate with the LEDs from a maximum distance of 480 cm. As our positioning approach requires measurement of the direct distance from the camera to the LED, we compare the accuracy of measuring the distance at different floor positions. Maintaining the floor height to be constant, we experience an increase in the percentage of error when the camera moves a long distance horizontally. To mitigate this error, we fix some recorded distance values within the application to compare. We have found maximum 2 cm error at 100 cm horizontal distance from the floor position of the target LED when the MR moves horizontally toward that LED. Thus, the accuracy is improved from the localization and navigation viewpoint as well as based on the distance for indoor communication.

## Conflicts of Interest

The authors declare that they have no conflicts of interest.

## Acknowledgement

"This research was supported by the MSIT (Ministry of Science and ICT), Korea, under the Global IT Talent support program (IITP-2017-0-01806) supervised by the IITP (Institute for Information and Communication Technology Promotion)"